\documentclass[manuscript,  nonacm]{acmart}
\usepackage{csquotes}

\usepackage{ifdraft}
\usepackage{etoolbox} 
\usepackage{xspace}
\usepackage{color}
\usepackage{tabularray}
\usepackage{comment}
\usepackage{cleveref}
\renewcommand*{\cref}{\Cref}
\usepackage{hyphenat}
\usepackage{array}
\usepackage{booktabs}
\usepackage{tabularx} 
\usepackage{svg}
\usepackage{subcaption}
\usepackage{longtable}

\usepackage[textsize=scriptsize]{todonotes}

\AtBeginDocument{%
  \providecommand\BibTeX{{%
    \normalfont B\kern-0.5em{\scshape i\kern-0.25em b}\kern-0.8em\TeX}}}

\NewDocumentCommand{\pquote}{+O{} +m}{\textit{\blockquote[#1]{#2}}}
\newcommand{\elide}[1]{\textelp{}} 
 
\begin{document}

\title{"I Talked to an AI Chatbot, So What's Next?" How U.S. Young Adults Imagine Responsible AI for Emotion Coping}

\author{Jiaying Liu}
\affiliation{%
  \institution{The University of Texas at Austin}
  \city{Austin}
  \country{USA}}
\email{jiayingliu@utexas.edu}

\author{Nimra Ishfaq}
\affiliation{%
  \institution{The University of Texas at Austin}
  \city{Austin}
  \country{USA}
}
\email{nimraishfaq@utexas.edu}
\renewcommand{\shortauthors}{}

\begin{abstract}
Emotion coping is inherently relational, unfolding through interactions with friends, family, professionals, and communities. Yet AI chatbots are largely designed around a user--AI dyad. Learning from the \textit{ethics of care}, we examine how AI chatbots shape the \textit{relational conditions} of emotion coping. We conducted a scenario-based study with 17 U.S. young adults across four emotion coping scenarios. Participants identified eight roles through which AI could support relational conditions, alongside three challenges: flattening distinct relational conditions, discouraging reciprocity, and shifting relational labor onto users. This study contributes a relational perspective of responsible AI in emotion coping. We argue that responsible AI should respond to individuals' situated relational conditions  rather than provide general-purpose support. We further identify two design principles: fostering \emph{reciprocity} by supplying materials to engage with others, and strengthening \emph{emotional self-efficacy}. Together, we position responsible AI as AI in the loop of human relationships.

\end{abstract}


\maketitle

\section{Introduction}
Emotion coping refers to the cognitive and behavioral efforts through which people manage, modulate, and make sense of emotionally demanding experiences \cite{grossEmotionRegulationCurrent2015,lazarus1984stress}. AI chatbots have increasingly become another source of support, valued for their constant availability, perceived low social risk, and low barriers to engagement \cite{murnanePersonalInformaticsInterpersonal2018}. However, emotion coping is inherently relational: people draw on friends, family, professionals, and communities for validation and pathways toward resolution \cite{thoits2011mechanisms,rickwoodYoungPeoplesHelpseeking2005,liuRegulationSupportCentering2025}.

Responsible design for emotion coping therefore requires considering not only how AI interacts with an individual user, but also how its involvement affects the relationships through which coping unfolds. Principles commonly emphasized in responsible AI, such as transparency, accountability, privacy, and fairness, are essential, but often focus on properties of the system and its interaction with individual users \cite{floridiAI4PeopleAnEthicalFramework2018,unescoRecommendationEthicsArtificial2022}. In the context of emotion coping, the \textbf{ethics of care} extends this perspective by treating people's interdependence, responsibilities toward others, and vulnerabilities within relationships as central to evaluating what constitutes responsible support \cite{gilliganDifferentVoicePsychological1993,noddingsCaringRelationalApproach2013}.

This relational perspective directs attention to the \textbf{relational conditions} under which AI enters, supports, and reshapes emotion coping, including how people connect with others, exchange care, and maintain their capacities to navigate emotional experiences \cite{vallorArtificialIntelligenceImperative2023}. Responsible AI for emotion coping thus requires examining not only what chatbots do for users, but also what their involvement does to the relationships around them. As part of a larger project investigating responsible AI for emotion
coping, this paper asks two research questions:

\begin{description}
  \item[RQ1:] What roles can AI chatbots play in shaping the relational
              conditions of emotion coping?
  \item[RQ2:] What challenges may arise as AI chatbots shape these
              relational conditions?
\end{description}



\section{Related Work}
In this section, we first situate emotion coping as a relational process, then review the growing landscape of AI chatbot support for emotion coping. Finally, drawing on the ethics of care, we develop a relational perspective on responsible AI that foregrounds how chatbots may shape the conditions through which people seek, give, and sustain care.

\subsection{Emotion Coping as a Relational Process}
Emotion coping refers broadly to the cognitive and behavioral efforts through which people manage emotionally demanding experiences, including how they regulate, interpret, and respond to emotions over time \cite{lazarus1984stress,grossEmotionRegulationCurrent2015}. Lazarus and Folkman conceptualize coping as efforts to manage demands that are appraised as taxing or exceeding one’s resources, while Gross’s process model describes emotion regulation through strategies such as situation selection, situation modification, attentional deployment, cognitive change, and response modulation \cite{lazarus1984stress,grossEmotionRegulationCurrent2015,kitsonSupportingCognitiveReappraisal2024}. Together, these perspectives establish coping as more than immediate reduction of negative affect: it involves making sense of emotional experiences, responding to the situations that produce them, and developing ways of managing subsequent demands.

Although foundational models often foreground processes within the individual, coping frequently unfolds through interactions with other people. Research on interpersonal emotion regulation shows that people routinely seek out others to influence their emotional states and likewise participate in regulating the emotions of those around them \cite{zakiInterpersonalEmotionRegulation2013}. Work on social sharing similarly demonstrates that emotional experiences are commonly processed through disclosure, conversation, and responses from others \cite{rimeIntrapersonalInterpersonalSocial2020}. Social-constructionist perspectives further emphasize that emotions are interpreted and expressed within social and cultural relationships rather than existing independently of them \cite{averillChapter12CONSTRUCTIVIST1980,harreSocialConstructionEmotions1986,barrettTheoryConstructedEmotion2017}. Across research on social support, disclosure, interpersonal conflict, and professional help-seeking, relationships therefore provide not only resources for coping but also contexts in which emotions are interpreted, negotiated, and acted upon.

These perspectives suggest that HCI research on emotion coping should account for the social relationships, responsibilities, and contexts through which people manage emotional experiences \cite{liuRegulationSupportCentering2025}. Doing so can broaden the design of emotion-support technologies beyond immediate individual regulation toward support for disclosure, interpersonal negotiation, sustained coping, and help-seeking.

\subsection{AI Chatbots for Emotion Coping}

AI chatbots increasingly participate in how people cope with emotionally difficult experiences. Research on conversational agents for mental health and wellbeing shows that users turn to them for accessible, non-judgmental spaces to disclose distress, seek validation, reflect on experiences, and receive guidance \cite{zouInfluenceIndividualsEmotional2026,kimUnveilingHumanTouch2025,mengExaminingContentForm2025}. In grief, for example, users have engaged chatbots as listeners, companions, emotion coaches, and simulations of deceased loved ones \cite{xygkouConversationLossUnderstanding2023,jiangReminiLeveragingChatbotMediated2025}. LLM-based systems further support journaling and reflection by helping users articulate and organize everyday experiences \cite{kimMindfulDiaryHarnessingLarge2024,nepalContextualAIJournaling2024}, while personalization through long-term memory can encourage continued self-disclosure and familiarity \cite{joUnderstandingImpactLongTerm2024,jiangScaffoldedVulnerabilityChatbotMediated2026,jiangRECALLbotDesigningAgentic2026}. More recent work on customizable LLM-based emotional-support agents similarly finds that users construct chatbots for emotional reliance, confronting stressors, self-reflection, and open disclosure \cite{zhengCustomizingEmotionalSupport2025,liCustomizableAIDepression2025}. Together, this work shows that chatbots are not used for a single form of emotional support: users move among venting, reflection, reappraisal, companionship, and practical problem-solving as part of everyday coping \cite{maBoundedSupportiveCompanionship2026}.

Much of this literature nevertheless takes the user--AI interaction as its primary unit of analysis, asking whether particular chatbot responses, features, or relationships improve disclosure, perceived support, engagement, or wellbeing. Emerging work has begun to complicate this dyadic framing. For example, research on grief considers whether chatbot use supports or inhibits renewed social connectedness \cite{xygkouConversationLossUnderstanding2023}; clinical systems such as MindfulDiary connect AI-mediated reflection to communication with mental-health professionals \cite{kimMindfulDiaryHarnessingLarge2024}; and recent studies of AI companionship examine withdrawal from offline relationships and broader psychosocial consequences
\cite{yuanMentalHealthImpacts2026}. Yet comparatively less attention has focused on how AI support becomes situated within users' existing relationships and whether repeated use strengthens, redirects, or displaces the capacities and human connections through which coping develops over time. This motivates our study to examine AI-supported emotion coping beyond the user--AI dyad.

\subsection{Ethics of Care and Responsible AI Design for Emotion Coping}

As AI becomes increasingly involved in mental health and emotional support, HCI research has raised concerns around transparency, privacy, fairness, accountability, safety, and appropriate reliance \cite{floridiAI4PeopleAnEthicalFramework2018,unescoRecommendationEthicsArtificial2022}. Work on mental-health technologies has similarly examined how sensitive data are collected and inferred \cite{asthanaKnowEvenIf2024}, how users understand system capabilities and limitations, and how designers can support informed and trustworthy use \cite{tavoryRegulatingAIMental2024}. These approaches provide important safeguards around the design of AI systems and users' interactions with them. Yet emotion coping also raises questions that are difficult to evaluate through properties of the system or outcomes for an individual user alone. AI chatbots' involvement may shape how users seek support, communicate with others, and distribute emotional work across their existing relationships. Recent HCI research on care similarly argues for understanding technology not only as a tool that delivers care, but as something that can become part of existing care relations \cite{wangCaringCareMetaNarrative2026}.

The \textbf{ethics of care} provides a foundation for examining these broader consequences. Rooted in feminist moral philosophy, care ethics foregrounds interdependence, vulnerability, attentiveness, and responsibility within relationships \cite{gilliganDifferentVoicePsychological1993,noddingsCaringRelationalApproach2013,vallorArtificialIntelligenceImperative2023}. Recent HCI scholarship has brought these commitments into design, emphasizing that ethical judgments are situated in relationships and social contexts rather than reducible to abstract principles or isolated individual outcomes \cite{henriquesFeministCareEthics2025,wangCaringCareMetaNarrative2026}. From this perspective, responsible AI for emotion coping concerns not only whether an interaction makes a user feel better, but also how AI participation shapes the \textbf{relational conditions} surrounding that interaction, such as people's connections to others, patterns of dependency, and capacities to navigate support.

HCI research on technologies for relatedness similarly suggests that technologies can be evaluated by how they support connections and capacities beyond engagement with the technology itself \cite{shiMappingCaregiverNeeds2026,vafafarWordsAreNot2026}. Building on these perspectives, we examine AI's involvement in emotion coping through the relational conditions it may create, support, or disrupt, with particular attention to whether AI chatbots foster users' capacities for independent coping and sustain connections with human support networks.



\section{Method}

\subsection{Scenarios-based design}

We adopted a scenario-based design approach \cite{carrollScenariobasedDesign1997, 
carrolFiveReasonsScenariobased1999}, in which concrete descriptions of use 
situations serve as the medium for reasoning about a system that does not yet 
exist. We used these scenarios to anchor design ideation in realistic situations reflecting common emotional scenarios young adults encounter and the support they seek in response. Because judgments about AI's appropriate role are highly context-dependent, this approach let participants reason about what responsible AI involvement would look like.

\subsubsection{Four Scenarios}
\label{sec:scenarios}

\begin{figure}[htbp!]
    \centering
    \includegraphics[width=0.8\textwidth]{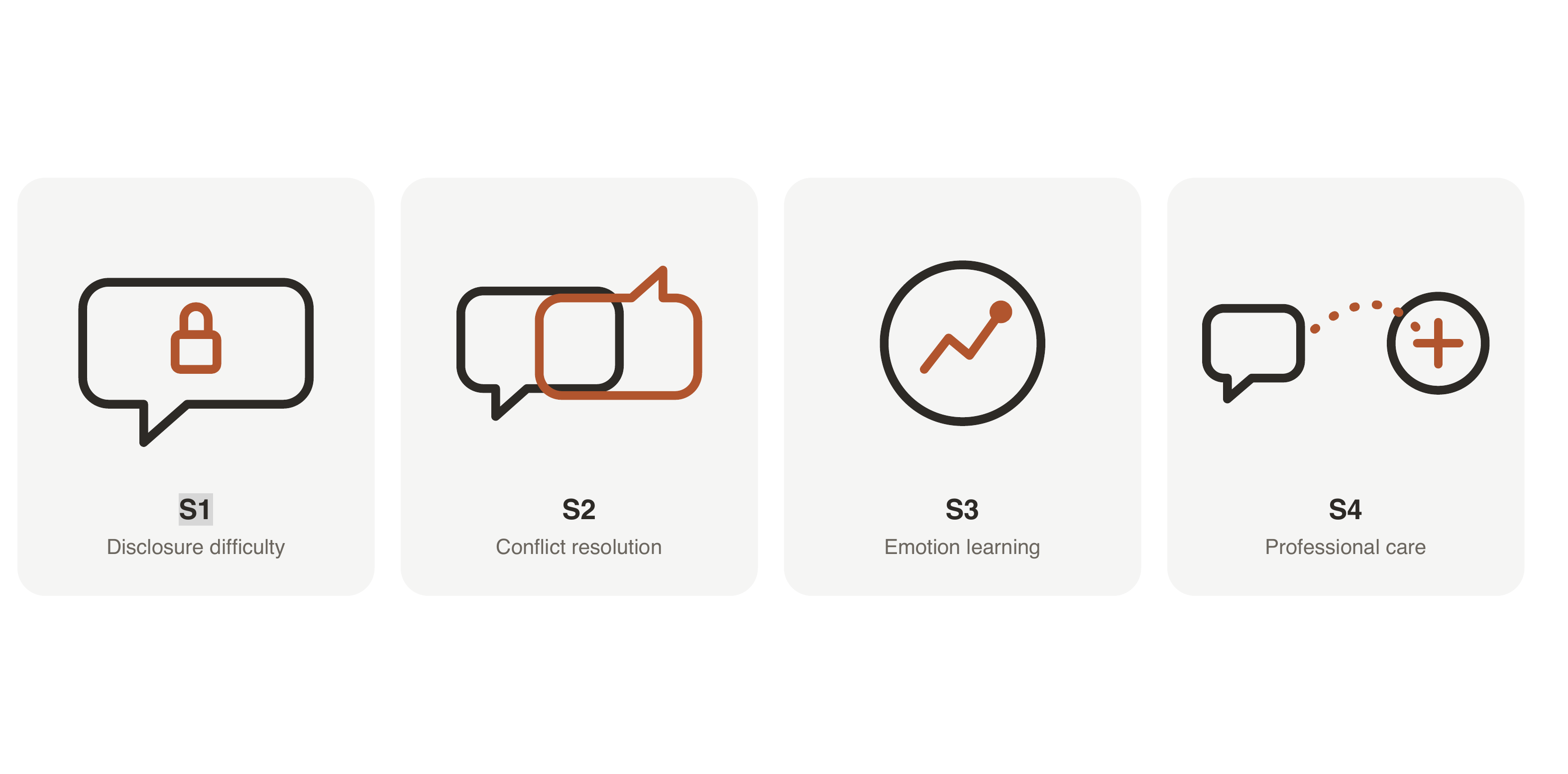}
    \caption{Four Scenarios}
    \label{fig:4scenarios}
\end{figure}

We selected the scenarios based on prior literature on young adults' emotional coping needs, choosing relational situations that are both frequently reported and persistently difficult to navigate (see Figure \ref{fig:4scenarios}). 
\textit{Disclosure difficulty} (S1) captures situations where fear of judgment or of burdening others keeps distress unspoken, and \textit{interpersonal conflict} (S2) demands emotion coping and perspective-taking in the moments. These two scenarios correspond to domains of the interpersonal competence framework \cite{buhrmester1988five}, which identifies self-disclosure and conflict management among the core relational capabilities people draw on to sustain close relationships---and which young adults commonly report struggling with. The other two scenarios extend beyond the immediate coping episode. \textit{Emotion learning} (S3) addresses the development of coping skills that shape how future distress is met, and \textit{bridging professional care} (S4) addresses barriers to help-seeking from formal care. 
Together, these scenarios span interpersonal, individual, and professional contexts. 
Table~\ref{tab:design_scenarios} lists each scenario alongside the related prior work.


\begin{table}[htbp!]
\caption{Design Scenarios}
\centering
\small
\begin{tabularx}{\textwidth}{p{1.5cm} p{10cm} p{2cm}}
\toprule
\textbf{Scenario} & \textbf{Scenario description} & \textbf{Previous study} \\
\midrule
\label{tab:design_scenarios}
 S1: Disclosure difficulty & You have been struggling with your mental health but feel afraid to tell someone close to you — a friend, family member, or partner. You worry about being judged, becoming a burden, or damaging the relationship. You are not sure whether to say anything, or how. You turn to an AI chatbot — not to be comforted in their place, but to work out whether, when, and how to say it. & \cite{Pri24,Wec24,Ras20,Fox21,Spe25,Whi13} \\
 S2: Conflict resolution & You are in the middle of a frustrating conflict with someone in your life — a parent, roommate, or partner. You feel dismissed, misunderstood, or overwhelmed by strong emotions. You are not sure how to respond or whether to reach out to them. You turn to an AI chatbot — not to take your side, but to help you steady yourself and decide how to respond. & \cite{Bau21,Bau24,Jun15,Yon24b,San07,Mcc24} \\
 S3: Emotion learning &  You have been thinking about how you deal with difficult emotions and would like to build skills to navigate them differently. You turn to an AI chatbot — not for comfort in the moment, but to design your own learning path, one that is intended to help you need it less over time. You can pick what feels right, mix approaches, or adjust how you use the AI as you grow. & \cite{Sha23,Smi22,Hou22,The22,Rot18,Kal19,Vau20} \\ 
 S4: Professional care & You have been thinking about seeking professional support for your mental health, or you are already seeing someone but find the time between appointments difficult to navigate. You are not sure how to find the right help, prepare for an appointment, or maintain progress between sessions. You turn to an AI chatbot — not as the therapist, but to help you find, prepare for, and stay engaged with human care. & \cite{Stu20,Rou25,Pfe22,Wil20,Kos23,habichtGenerativeAIEnabled2025,Cam25,Cao25b} \\
\bottomrule
\end{tabularx}
\end{table}

\subsubsection{Design Materials}
\begin{figure}[htbp]
    \centering
    \includegraphics[width=0.95\textwidth, alt={Examples of two completed Miro boards. Participants dragged the stickers that they want to the corresponding areas described in 3.2.}]{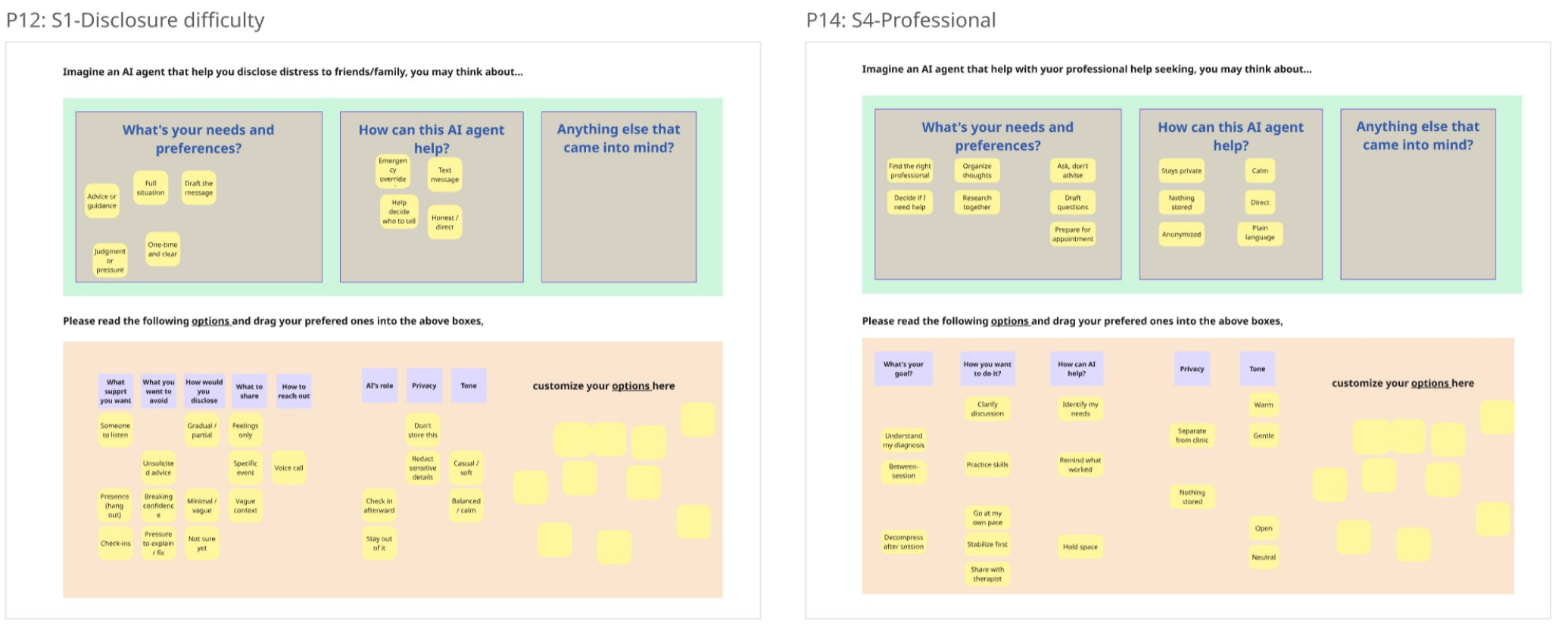} 
    \caption{Examples of two completed Miro boards.}
    \label{fig:miro}
\end{figure}

For each scenario, we developed a structured Miro board as a workspace 
for design ideation. 
Each board ran two parallel tracks distinguishing what the \textit{human support network} should do from what the \textit{AI chatbots} should do, making the division of labor an explicit design decision and eliciting participants' intuitions about what AI should do, how it should act, and where its involvement should give way to human support. The examples of two completed Miro boards are showed in Figure \ref{fig:miro}.

\subsection{Procedure}
All sessions were conducted over Zoom and lasted approximately 60 minutes. Each participant was compensated with a \$20 Amazon gift card.
The first phase, approximately 20 minutes, introduced participants to the scope of the study and asked about their current emotion coping strategies and use of AI chatbots, establishing a baseline understanding of their existing habits, expectations, and points of friction. Participants were then introduced to the four scenarios. 

The remaining 40 minutes were dedicated to the scenario-based design activity. 
Participants selected two of the four scenarios that they felt were most relevant to their own experiences. This selection was intentionally participant-driven, ensuring that design ideation remained grounded in personally meaningful cases rather than hypothetical contexts. For each selected scenario, the interviewer walked participants through the corresponding Miro board, and participants spent around five minutes arranging the design cards to reflect their imagined AI chatbots in that scenario. They could modify, reject, or add cards as needed. The interviewer then used the emerging board as a probe for discussion, asking participants to explain their choices, elaborate on the boundaries they drew between the human support network and AI, and reflect on what conditions would make them more or less comfortable with AI chatbots involvement. 

\subsection{Recruitment and Participants}

Recruitment for this study utilized a southern U.S. university listserv and Reddit platforms to attract young adults from diverse backgrounds. Given Reddit’s effectiveness in reaching historically underserved communities, recruitment posts were placed in regional marketplace subreddits such as r/SanAntonioJobs and r/NewOrleansMarketplace.
Interested participants aged 18--25 completed a screening questionnaire that included demographic information, PHQ-4, and AI chatbot interactions. PHQ-4 is a widely used four-question tool which captures anxiety and depressive symptoms over the previous two weeks \cite{kroenkeUltraBriefScreeningScale2009}. Participants with PHQ-4 scores indicating mild to moderate symptoms were selected for ethical reasons to ensure meaningful emotional distress experiences while avoiding those needing immediate clinical intervention. 

We reviewed around 150 screening survey responses. A theoretical sampling strategy was employed, guiding participant selection iteratively through data analysis over three rounds of data collection from April to July 2026. This involved purposive sampling to maximize variation in demographic characteristics and mental health status. For example, the initial round’s findings revealed the need for more critical views on AI chatbots and participants with formal help-seeking experiences, prompting targeted recruitment adjustments.

\newcolumntype{Y}[1]{>{\RaggedRight\arraybackslash\hsize=#1\hsize}X}

\begin{table*}[htbp!]
\centering
\scriptsize
\renewcommand{\arraystretch}{1.15}

\caption{Participants}
\label{tab:phase2-participants}

\begin{tabularx}{\textwidth}{
    >{\hsize=0.3\hsize}X  
    >{\hsize=0.3\hsize}X  
    >{\hsize=0.7\hsize}X  
    >{\hsize=1.6\hsize}X  
    >{\hsize=1.1\hsize}X  
    >{\hsize=1.6\hsize}X  
    >{\hsize=1.1\hsize}X  
    >{\hsize=1.1\hsize}X  
    >{\hsize=1.2\hsize}X  
}
\toprule

\textbf{ID} &
\textbf{Age} &
\textbf{Gender} &
\textbf{Education} &
\textbf{Race} &
\textbf{Background} &
\textbf{Diagnosis} &
\textbf{PHQ-4} &
\textbf{AI Use} \\

\midrule

P01 & 25 & Female & Graduate degree & White & LGBTQ+ & Social anxiety & Mild & Occasionally \\

P02 & 22 & Male & Some college & Black & -- & Depression & Mild & Frequently \\

P03 & 23 & Male & High school & Black & LGBTQ+ & No & Moderate & Frequently \\

P04 & 25 & Female & Graduate degree & Asian & First-gen; Low income & No & Moderate & Occasionally \\

P05 & 22 & Male & Some college & Black & First-gen & Depression & Moderate & Frequently \\

P06 & 25 & Female & Graduate degree & Asian & International student & No & Mild & Occasionally \\

P07 & 24 & Female & Some college &  Black & -- & Depression & Moderate & Occasionally \\

P08 & 25 & Male & Graduate degree & Asian & International student & No & Moderate & Frequently \\

P09 & 25 & Male & Some college & Asian & -- & No & Moderate & No \\

P10 & 24 & Female & College graduate & Asian & -- & No & Moderate & Once or Twice \\

P11 & 20 & Male & High School & Black & Low income & No & Moderate & Occasionally \\

P12 & 24 & Female & College graduate & Black & Low income & Depression & Moderate & Frequently \\
P13 & 22 & Female & Some college & Black &  & Anxiety & Mild & Occasionally \\
P14 & 23 & Female & College graduate & Asian &  & No & Moderate & Once or twice \\
P15 & 23 & Male & College graduate & Black  & International students & No & Moderate & Occasionally \\
P16 & 24 & Male & High School & White & First-gen & Depression & Mild & Frequently \\
P17 & 25 & Male & College graduate & Black & First-gen & No & Moderate & Frequently \\
\bottomrule
\end{tabularx}

\footnotesize
Notes: PHQ-4 severity categories were coded as Mild (3--5), Moderate (6--8), and Severe (9--12). 

\end{table*}

\subsection{Data Analysis}
All sessions were audio-recorded and transcribed by Zoom. Immediately following each session, the first author wrote a debrief memo capturing initial impressions, emergent questions, and observations about the participant's engagement. Transcripts, memos, and debriefs were imported into NVivo for analysis.
Analysis proceeded through iterative open and axial coding \cite{strauss1990basics}. During open coding, the authors read transcripts line by line, assigning descriptive codes close to participants' own language. 
Axial coding then related these open codes, specified the conditions under which patterns held, and grouped codes into provisional categories. 
The codebook was revised across four iterations, consistent with the constant 
comparative logic of theoretical sampling \cite{charmaz2007grounded}, in which 
subsequent interviews developed and saturated emerging categories.

\subsection{Ethical Considerations}

Given the sensitive nature and psychological demands of this study, we implemented several measures to protect participants' privacy and comfort and to encourage open disclosure during the interview. At the beginning of each session, participants were assured of the non-judgmental and confidential nature of the study; except for the email addresses, no names or other identifiable information  were collected at any point. Participants were also informed of their right to decline any question they found uncomfortable and to withdraw from the study at any time.

\section{Findings}
In Section 4, we map the roles chatbots could play across the four emotion coping scenarios as they emerged from the scenario-based design activities (RQ1).

\subsection{RQ1: The Roles of AI in Shaping Relational Conditions}

The four design scenarios placed emotion coping in distinct relational situations: disclosing vulnerability, navigating conflict, developing capacities for future interactions, and engaging with professional care. Across these contexts, participants imagined AI not simply as a source of direct emotional support, but as taking roles that could shape the conditions under which people approached, participated in, and sustained relationships involved in coping. We identified eight such roles, presented by scenario below. Section 6 further considers the principles that cut across these roles.

\begin{figure}[htbp]
    \centering
    \includegraphics[width=.9\textwidth]{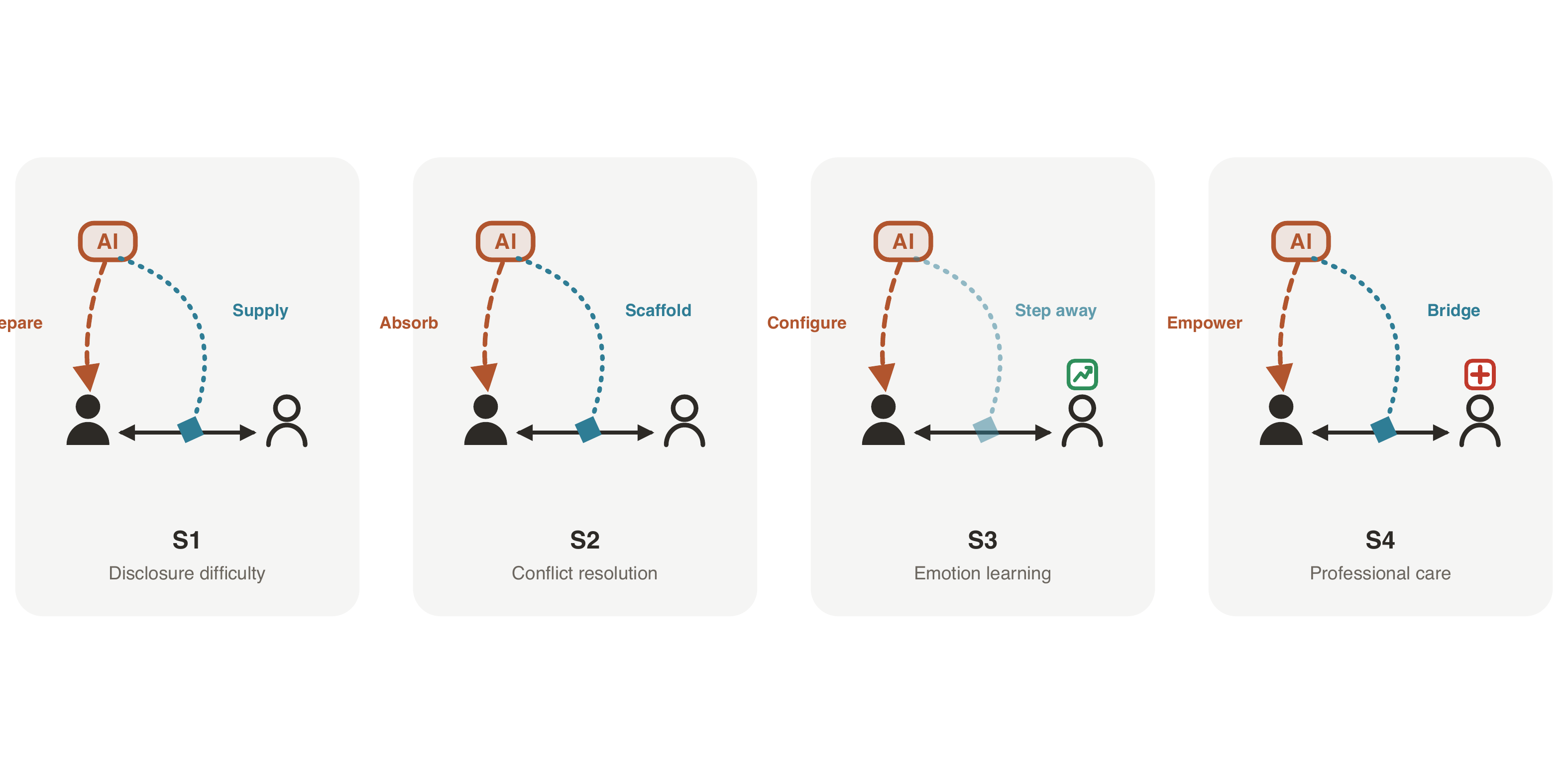}
    \caption{Roles AI chatbots can play across the four design scenarios.}
    \label{fig:roles}
\end{figure}

\subsubsection{S1: Supporting the Conditions for Disclosure --- Prepare and Supply}

Disclosure places individuals in the vulnerable position of deciding whether, how, and how much to reveal to another person. Participants imagined AI helping them become ready to speak and making what they wanted to communicate easier for others to understand.

\textbf{Prepare individuals to enter disclosure with clarity, confidence, and boundaries.}
Before approaching another person, participants used AI to work through what they were feeling, decide what they wanted to say, and anticipate how the conversation might unfold. Some first used the chatbot as a low-stakes space for turning an unstructured stream of thoughts into a clearer account. Others emphasized rehearsal, particularly when disclosure involved someone whose reactions felt difficult to predict. P09, for example, supplied details about how the other person typically behaved so that AI could help plan both wording and boundaries: \pquote{I asked AI...this is the kind of discussion I need to have... And this is usually how things go whenever I talk about important things with this person... how to approach it, the verbiage that I could use to not trigger them... and if need be, how to disengage.} Here, preparation was not only about composing a message, but about entering a potentially vulnerable interaction with greater clarity over what to say and how far to engage.


\textbf{Supply individuals with materials that help others understand and recognize their experience.}
Participants also wanted something they could carry into the disclosure itself. This ranged from language for articulating an experience to external information that could make it more recognizable to others. P10, for instance, asked AI for evidence that could help explain her feelings to family members: \pquote{AI really helped me validate my feelings on this. And then I was able to use those facts... so I could explain to my family members... What I'm going through is legitimate, and I'm not making up anything.} Others used AI to turn difficult-to-express thoughts into outlines or draft language. For example, P13 described AI as helping produce an outline: \pquote{I think chat has been very helpful with helping me figure out what to say. And it'll give me an outline.} Rather than speaking on the user's behalf, participants wanted AI to supply resources that could travel with them into the human conversation.


\subsubsection{S2: Supporting the Conditions for Constructive Conflict --- Absorb and Scaffold}

Conflict differs from disclosure in that individuals are already situated within a strained interaction, where heightened emotion and poorly framed communication can further affect the relationship. Participants therefore positioned AI around the interpersonal exchange: first helping contain reactions that could intensify the dispute, and then helping structure communication.

\textbf{Absorb aggression and self-doubt before they spill into the interpersonal exchange.}
Before confronting another person, participants used AI to work through heightened emotions and uncertainty about their own position. Some wanted concrete techniques for regulating stress before it turned into aggression. 
P12 emphasized physiological escalation:\pquote{When you're stressed, [...] you might not be able to handle your emotions at that moment, and you will be passing aggression.} Others treated AI as a comparatively detached audience for reconsidering whether they themselves had contributed to the conflict. P05 described asking the chatbot to assess an argument and accepting its judgment that he had crossed a line: \pquote{AI told me I was the one who was on the wrong side. And I had to apologize.} At other times, agreement from the AI strengthened his conviction. In this role, AI became an intermediate space in which immediate reactions could be processed before they entered the relationship.

\textbf{Scaffold communication while preserving the individual's own voice and relational agency.}
Once participants were ready to engage, they wanted help finding language and structure for difficult messages. They described treating AI suggestions as material to adapt. Many explicitly expressed aversion to directly copying and pasting the AI-generated message to transmit verbatim. P09 emphasized retaining ownership of the final wording: \pquote{If it says anything that I like, then I'll take that and incorporate it into my own message, into my own words. So that whatever I say is still my words.}
More interestingly, some saw this restructuring as changing the interaction itself. P09 described AI helping externalize a contentious issue so that it became something both people could examine rather than an accusation directed at one person: \pquote{It was taking the frame of the idea of what you're saying off of the person, and kind of laying it out in front of both of you. So that it doesn't feel like the other person is in the spotlight, or being targeted}. Participants also reported learning communication structures from repeated use, such as cushioning disagreement or beginning with their own feelings. For example, P04 described learning a structure for communicating disagreement with family: \pquote{I kind of learned from that. I didn't copy it, but the structure of it... I'm kind of learning the structure by heart.}

\subsubsection{S3: Building Capacity for Relational Emotion Coping --- Configure and Step Away}

Emotion learning was not organized around a single immediate interaction, but around developing capacities that could carry across future emotional and relational situations. Participants imagined support that helped cultivate those capacities and gradually became less necessary as they developed.

\textbf{Configure learning that builds individuals' capacity to cope within relationships.}
Participants wanted AI to organize reflection, practice, feedback, and habit formation into a learning process rather than offer isolated coping tips. They envisioned that AI chatbots could configure structured exercises, tracking existing habits, role-play, and situations that could reveal weaknesses that were difficult to identify through self-report alone. P12 elaborated that, \pquote{I'd love to have some offline materials, role play, and daily habits - reflection as well... give feedback, because it helps you grow.} P10 proposed having AI generate scenarios and evaluate how she responded: \pquote{I would have AI kind of create situations, case scenarios, and then see what kind of responses I would give it. Because sometimes, you know, you don't know what you are not good at.} Such configurations positioned AI as a temporary learning environment in which people could develop skills intended for use beyond the chatbot itself.


\textbf{Step away at a personalized speed and stop responding if necessary.}
As those skills became more established, participants expected the AI to recognize their progress and reduce its involvement. They wanted suggestions to reflect what they had already learned or tried, with familiar problems gradually requiring less intervention. P06 imagined the system beginning to withdraw when repeated situations indicated increasing competence, while P09 described the desired trajectory more explicitly as developing \pquote{the use of skills outside of AI, trying to be not fully reliant on AI}, followed by support that would \pquote{then gradually decrease...you take the training wheels off. Then it can slowly just step away and be like, `Okay, you got this.'} The endpoint of the role was therefore the reduced need for AI chatbots. While most participants preferred the process of stepping away to be gradual, one participant instead argued that AI chatbots could be more radical when they step away. P13 said, \pquote{I think maybe AI should stop responding to me it when sees the same question occur again and again}.


\subsubsection{S4: Supporting the Conditions for Professional Care --- Empower and Bridge}

Professional care introduced a different relational setting, shaped by limited communication with professionals and practical work that individuals must navigate before and between appointments. Participants saw AI as useful both for helping them take a more active role in treatment and for reducing barriers surrounding access and continuity.

\textbf{Empower individuals through early pattern recognition and greater ownership of their treatment plan.} Participants wanted AI to help individuals recognize patterns earlier, make sense of clinical interpretations, and take a more informed role in shaping their treatment.
First, participants valued AI's ability to surface recurring patterns in their emotion-coping experiences before those patterns developed into concerns requiring clinical attention. P05, for example, described how the chatbot's memory could reveal repeated difficulties: \pquote{if there is a lot of repetition of similar problems that I talked to AI [...] it can tell me that you have been thinking like this for a couple of times, things like that.} In addition, entering professional care also involves learning to connect one's lived experiences with clinical concepts and terminology. Participants saw AI as potentially making this interpretive process more continuous and accessible. P06, for instance, wanted AI to help her understand a diagnosis that had initially been difficult to recognize in her own experience: \pquote{It was only through therapy that I could understand that it was anxiety, or depression. So, if AI could help me understand the diagnosis given by the therapist earlier, that would also be helpful.} Participants further extended this support toward greater ownership of the treatment process itself. P01 imagined using AI to think through the therapeutic approach and topics she wanted to bring into a session with the therapist: \pquote{Specifically for me, being able to choose the approach - most of the time, it's going to be Cognitive Behavioral Therapy - being able to specify an approach, specify an overall topic that you want to come back to}. 


\textbf{Bridge individuals to professional care by improving access and continuity.}
Participants also imagined AI supporting the practical work around the clinical encounter. Finding an appropriate professional could itself require substantial effort: P06 described manually reviewing therapists and wanted AI to assist with that screening process: \pquote{Finding the right professional was very difficult. I had to manually sit through, check their resume and stuff like that. So if [the AI] can help me with it, it would be quite nice.} Once care had begun, participants wanted similar support preparing for appointments and organizing what had happened between sessions (P12). In this role, AI remained outside the clinical relationship while helping people reach it and make better use of the limited time.


\section{Challenges of AI chatbots shaping relational conditions for emotion coping}

We examined the challenges that may arise as AI chatbots enter and shape the relational conditions of emotion coping. Participants described several ways in which chatbots may neglect these conditions or fall short.

\subsection{AI chatbots can flatten distinct relational conditions}

Participants described that current AI chatbots can struggle to preserve these distinctions, instead flattening diverse relational conditions through a relatively uniform conversational interface and mode of support, while also remaining insufficiently attentive to nuanced relational privacy needs.


\textbf{Relational differentiation} concerned whether the chatbot recognized that different relationships called for different forms of support. Participants often had to reconstruct these distinctions themselves. P08 wanted persistent understanding of the particular person he was discussing: \pquote{It'll be nice to have a persona about the person that I'm talking about... so I don't have to go through and give enough context that 'Hey, the person's going through (this)'. It already knows they're going through a bad situation in life.} P03 similarly created a separate therapist persona to keep professional support distinct from other conversations: \pquote{On the therapist chat character, I'll only stick to professional conversations.} In both cases, participants themselves supplied distinctions---between people and between kinds of care---that the chatbot did not reliably maintain.

\textbf{Relational perspective} concerned whether the chatbot preserved the multiple perspectives that may constitute a relationship. In interpersonal conflict, participants worried that a chatbot could act confidently on one person's account of a situation involving others. P15 wanted the chatbot to resist turning the absent person into an antagonist: \pquote{One thing I want is for AI not to make the other person feel like a villain. Conversations are complicated, so I wouldn't want you to just agree with me and say, okay, you are right, and they are wrong. It should help me understand different perspectives}. P08 similarly objected to the chatbot imposing a definitive interpretation: \pquote{What it absolutely shouldn't do is make the problem worse by saying, oh, this is life-threatening, this is a deal-breaker, this is that and this. Like, let me decide that}. P14 instead wanted AI chatbots to preserve interpretive space: \pquote{It would be helpful to either give multiple interpretations, to sort of allow me to make a decision. Or asking a question, like, what do you think?}

\textbf{Relational privacy} concerned two forms of exposure created through emotionally sensitive chatbot conversations: who around the user might see the conversation through the interface, and how information about other people discussed in the conversation was represented and protected. P08 described routinely hiding the sidebar after disclosing personal distress because he was \pquote{scared of people looking at it}, showing how persistent histories and visible interface elements could expose private coping to others in the user's physical environment. Privacy concerns also extended to the people being discussed. P10 described deliberately removing identifying information when seeking advice about interpersonal situations: \pquote{If you are having to go to AI to get a solution for something, you can try to generalize it, or just don't put names onto who you're dealing with, and just put person A and person B for privacy's sake}. These accounts show that the need for relational privacy involves both protecting the user's conversation from unintended viewers and protecting the identities and information of other people brought into that conversation.

\subsection{AI chatbots can short-circuit reciprocity with others}

A second challenge concerned whether emotion coping remained embedded in reciprocal relationships with other people. Participants' accounts suggested that chatbot support could weaken this relational condition by moving two-way exchanges into a private user--AI interaction, where other people had fewer opportunities to jointly contribute their own perspectives, respond to the user's understanding of a situation, or remain involved in an ongoing process of sense-making. 

This was especially salient in interpersonal conflict, where the chatbot typically encountered only the user's account and could reinforce that interpretation without exposing it to another person's perspective or response. P14 described this informational asymmetry: \pquote{when people are speaking to AI, they're giving a small snapshot of what's going on with them}. What was missing, however, was not simply additional context, but the reciprocal process through which people explain themselves, encounter perspectives not already organized around their own account, and potentially revise their understanding. P05, for example, described using the chatbot's agreement to reinforce his position before confronting someone else: \pquote{ChatGPT agrees, and I feel more confident. When facing something, I know I'm on the right. I tell them (the third party) we can ask (the chatbot)}. P14 similarly worried that \pquote{AI can have bias in your favor...}. In these cases, AI could become another source of confirmation within one person's interpretation rather than supporting reciprocal sense-making with the other people involved.

A similar concern arose when emotional processing with AI became disconnected from ongoing relationships of care. This was particularly consequential in professional support, where participants could reach decisions with AI that were not visible to their clinicians. P13, for example, used a chatbot to research her medication and later stopped taking it, while her doctor did not know how she had reached that decision: \pquote{I told them I was going to do research on it. I don't think I told them how I did the research}. She explained, \pquote{they gave me medicine, but I stopped taking it, just because I don't feel like it was working, and it was making me lose my appetite}. Here, participants' reasoning process had become detached from communication with their clinicians, limiting opportunities for clinicians to respond, clarify, or incorporate that reasoning into subsequent care.

Several participants also recognized this gap and imagined ways for AI to reconnect these relational exchanges. P13 proposed a consent-based mechanism for bringing AI-mediated conversations back into professional care: \pquote{I think that it's okay if there's a choice if you are deciding you consent to ChatGPT (for) sharing this particular conversation with your therapist.} P04, P16 and P17 also had similar ideas around sharing their chat history with clinicians through user-controlled exports.

\subsection{AI chatbots can shift relational labor onto users, demanding emotional self-efficacy}

The preceding challenges also had a consequence for users' own role in sustaining appropriate relational conditions. When chatbots flattened distinctions among relational needs or lacked the reciprocal perspectives through which misunderstandings could be challenged, users need to compensate by identifying their own needs, articulating boundaries, judging responses, and correcting the interaction. In this sense, chatbot support could shift relational labor onto users and make \emph{emotional self-efficacy}---the capacity to understand and manage one's emotional needs and responses---a prerequisite for obtaining useful support.

Participants described performing this work even before substantive emotional support could begin. P10 had to anticipate potentially harmful responses and translate her own vulnerabilities into explicit instructions before discussing the problem itself: \pquote{I feel like I've had to train AI a lot when it comes to dealing with me. Before I told AI the entire situation, I told AI that I don't want to hear the generic kind of solutions. I told it that getting space is hard for me in this moment, so if you suggest that it will trigger me more.} Using the chatbot effectively therefore already required her to recognize her triggers, anticipate what kinds of advice might worsen her emotions, and articulate those limits clearly.

This evaluative work continued once the interaction was underway. P06 described the chatbot picking up her emotional tone and pushing it toward more extreme conclusions: \pquote{It picks up on the tone that I would come with. It would just give the most extreme answer, either yes or a no. And then I have to sit and sort of argue with it, with the context.} She therefore had to evaluate the response while distressed and supply the contextual moderation that she felt the chatbot had missed. 

Participants differed in how readily they could identify what support they needed, communicate those needs, and judge when the chatbot's responses were misaligned. Some therefore sought other ways to compensate. P01, for example, relied on someone else's pre-written personalization prompt: \pquote{It was a prompt that was really well created by somebody---that you could plug into ChatGPT.} This points to an important tension: effective chatbot support can require substantial emotional self-efficacy from users, even though responsible support might instead seek to strengthen that capacity.


\section{Discussion}
This study learned from the ethics of care and investigated how to support the relational conditions of emotion coping. Building on these findings, in this section, we first discuss \textbf{what responsible AI means} in the context of emotion coping, focusing on the situated relational conditions that AI should recognize. We then translate these principles into concrete \textbf{design implications} for AI chatbots.

\subsection{Designing Responsible AI Around Situated Relational Needs}

Work on responsible AI for mental health and emotional wellbeing has largely extended general principles of AI ethics, such as transparency, accountability, privacy, and fairness, into this domain \cite{floridiAI4PeopleAnEthicalFramework2018, unescoRecommendationEthicsArtificial2022, asmanResponsibleDesignIntegration2025}. Our study takes a different point of departure. Grounded in the \textbf{\textit{ethics of care}}, we treat the relational conditions of emotion coping as a central concern of responsible AI design. Responsibility, from this perspective, concerns not only how an AI system treats an individual user, but also how its involvement shapes the relationships through which care and coping take place \cite{gilliganDifferentVoicePsychological1993, noddingsCaringRelationalApproach2013, vallorArtificialIntelligenceImperative2023}.
From this perspective, responsible AI for emotion coping means \textbf{designing AI in the loop of human relationships} to support the people and relational processes through which coping unfolds. Unlike domains where AI is used primarily to automate work while retaining human oversight, emotion coping unfolds through relational conditions in which other people often remain integral to care.

We found that responsible AI cannot be organized around a uniform notion of ``emotional support.'' Instead, responsible AI must account for how relational needs vary across coping situations and calibrate AI support to these differences. Our scenario-based approach made this variation visible by grounding participants in four common and challenging situations: disclosing distress to others, navigating interpersonal conflict, developing lasting coping skills, and reaching out for professional help \cite{sandersProbesToolkitsPrototypes2014}. Across these scenarios, relational needs differed in their goals, stakes, pacing, and the people involved. However, findings suggested that the current AI chatbots seem to flatten individuals' diverse relational needs. For example, support that is appropriate when preparing to disclose distress may be inappropriate in a conflict where another person's perspective remains contested.

\begin{figure}[hbtp!]
    \centering
    \includegraphics[width=0.8\linewidth, alt={An illustration of the two principles of responsible AI for emotion coping- afford materials for reciprocity and emotional self-efficacy. They are described in detail in Section 5.2.}]{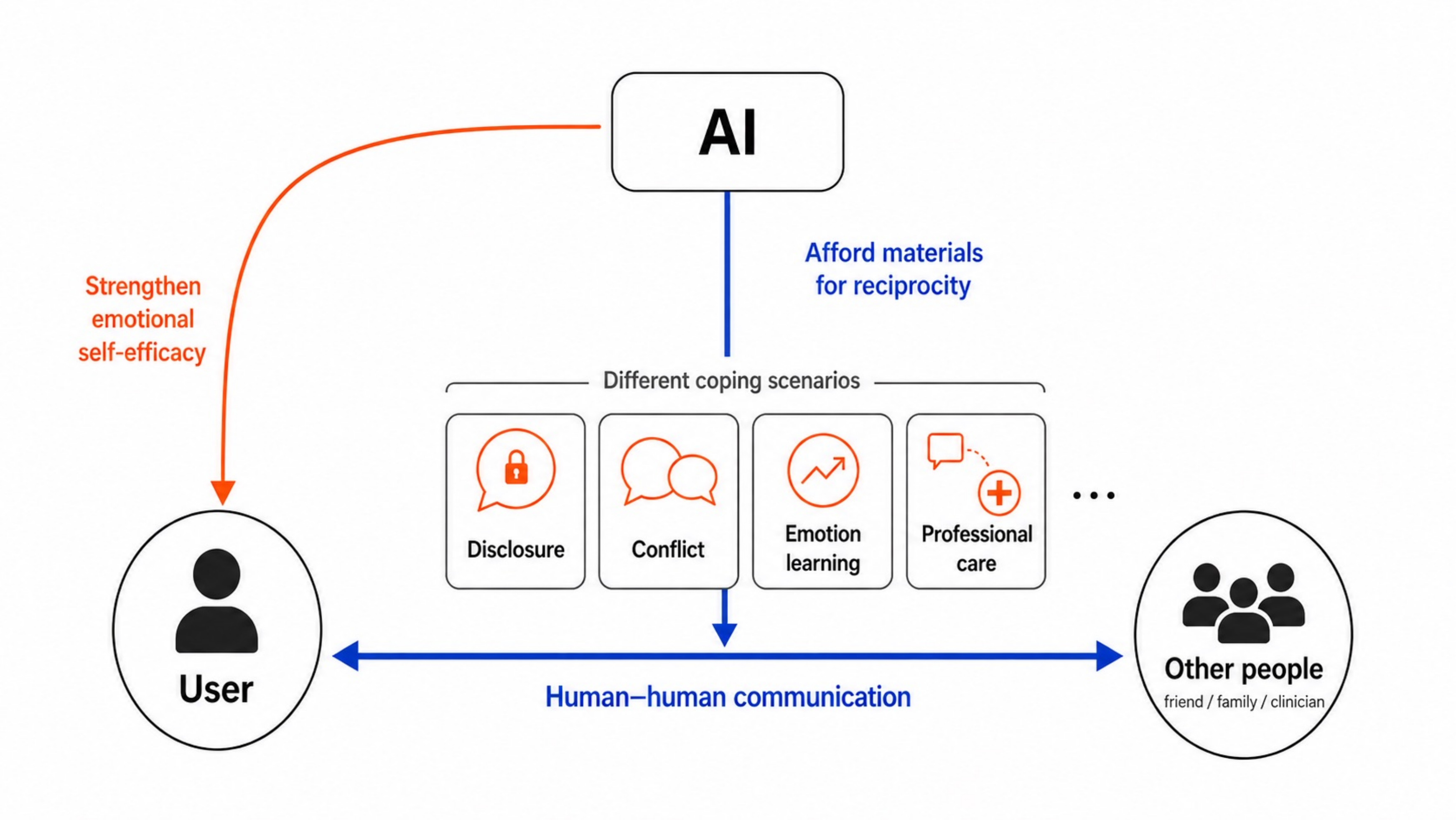}
    \caption{Two principles of responsible AI for emotion coping}
    \label{fig:mechanism}
\end{figure}

Specifically, we found two complementary ways AI can support the relational conditions of emotion coping (Figure \ref{fig:mechanism}). Some roles supported \emph{reciprocity} by helping users engage more effectively with other people; others strengthened \emph{emotional self-efficacy} by building knowledge, skills, and capacity that could persist beyond the immediate AI interaction \cite{banduraSelfefficacy1997, zouInfluenceIndividualsEmotional2026}. These two principles specify how responsible AI can support emotion coping without becoming its endpoint.

\subsection{Design Implications for Supporting Emotion Coping in Relational Contexts}
Extending the discussion before, we outline four specific design implications that directly inform clinicians, researchers, developers, and policymakers to help them design AI systems that promote reciprocity and emotional self-efficacy.

\textbf{Segment and customize support for diverse relational contexts.}
AI systems should help users separate different relationships, topics, and forms of support rather than treating emotion coping as one continuous conversational context. Existing chatbots provide one interface with continuous threads that burden the user with the tasks of providing context, organizing information, and managing privacy concerns \cite{10.1145/3757430}. 
Future systems could provide explicit entry and exit points for different relational contexts, along with granular controls over what information is remembered, shared, or visible. Binary toggles making conversations 'hidden' or 'visible' might be inadequate. Systems should let users define sensitive information, selectively reuse context, and customize memory across devices and settings \cite{song2026don, chen2025clear, mandal2025towards}. 


\textbf{Help users find their own words for human interactions.}
AI should help users articulate what they want to communicate, rather than speak on their behalf. Our participants described four ways this could happen: gathering evidence to legitimize their experiences, generating phrasing as raw material, calibrating the level of disclosure, and preparing briefs for clinical appointments. 
Thus, it is important to treat AI-generated language as revisable scaffolding for users' own expression rather than as finished communication. During drafting, AI could flag missing context, avoid exaggerating information or emotional cues, and offer multiple ways of expressing the same underlying intent \cite{tsengChatShouldLeave2026, wu2026caught}. In interpersonal conflicts, it could also ask about the other person's prior behavior and relational boundaries that should shape what is said. More broadly, systems should preserve the user's voice, defer to their preferences about whether and how to contact third parties, and provide ways for recipients to understand the provenance and extent of human contribution when appropriate \cite{fan2026still}.

\textbf{Make harmful agreement visible.}
Emotion coping support can become harmful when AI simply validates the user's interpretation rather than helping them examine it. Our participants often had to detect and correct this themselves: they objected when chatbots made unfair judgments about third parties, clarified emotional triggers, supplied missing context while distressed, and borrowed pre-written prompts to obtain better-calibrated responses. 
AI could surface uncertainty, distinguish validation of emotions from endorsement of interpretations, and offer alternative perspectives when appropriate. Systems could also ask about users' goals, values, and relevant relational boundaries before giving advice, while using explainability to make possible failures visible and correctable \cite{herrera2026co, wen2026made}. When the system cannot provide support without reinforcing harmful patterns, it should reduce its role rather than continuing the same form of engagement.

\textbf{Help users carry AI-supported coping into professional care.}
AI can help users carry insights developed during everyday emotion coping into interactions with professionals, rather than allowing those conversations to remain isolated within the chatbot. Systems could let users choose which summaries, patterns, questions, or excerpts to share and with whom. This flexibility is important because preferences for disclosure differ: some people may feel more comfortable first discussing sensitive issues with AI, while others prefer direct human support \cite{lee2020hear}. Clinicians should be able to engage with these materials without stigmatizing patients' AI use, particularly as prior work suggests greater acceptance of AI for lower-stakes activities such as documentation and information support than for complex clinical decision-making \cite{purohit2026conditional, thoits2011mechanisms}.

\subsection{Limitations and Future Work}

Our account of responsible AI focuses on the relational conditions of emotion coping and, specifically, on reciprocity and emotional self-efficacy. These concepts are not a complete framework for responsible AI. They complement concerns such as safety, privacy, fairness, accessibility, factual reliability, clinical efficacy, and responsibility allocation \cite{cooper2026framing, floridiAI4PeopleAnEthicalFramework2018, unescoRecommendationEthicsArtificial2022}. Future work should examine how relational design principles interact with these established dimensions of responsible AI.


\section{Conclusion}
This study shows that responsible AI support for emotion coping cannot be understood through the user--AI interaction alone. Across our scenarios, participants envisioned AI taking multiple roles that could reshape the \textbf{relational conditions} surrounding emotion coping, while also raising challenges around flattening relational needs, discouraging reciprocity, and shifting emotional labor onto users. These findings suggest that responsible design should consider how AI support affects people's relationships with others and their own capacity to cope. More broadly, this relational perspective offers a way to evaluate responsible AI by the conditions it creates around people.

\section{Acknowledgements}
We appreciated the valuable input from participants, sharing their emotionally vulnerable experiences and exploring how AI chatbots can support these experiences more responsibly. 
We used AI to help revise the text and generate figures under the guidance and close examination of authors. 

\bibliographystyle{ACM-Reference-Format}
\bibliography{MentalHealthproject, NimraLit, previous_work_scenarios}


\appendix

\end{document}